\documentclass[iop,apj,tighten,numberedappendix,twocolappendix]{emulateapj}
\usepackage[breaklinks,colorlinks,urlcolor=blue,citecolor=blue,linkcolor=blue]{hyperref}

\usepackage{amsmath,amssymb}
\usepackage{cleveref}
\usepackage{graphicx}
\usepackage{bm}
\usepackage{xcolor,ulem}
\usepackage[toc,title,page]{appendix}
\usepackage{tocbibind}
\usepackage{color}
\newcommand{\Msun}{{\rm M}_\odot}

\shorttitle{Persistent emission from sBHs in AGNs}
\shortauthors{et al.}

\begin{document}
\title{
High-energy electromagnetic, neutrino, and cosmic-ray emission by 
stellar-mass black holes in disks of active galactic nuclei
}

\author{
Hiromichi Tagawa\altaffilmark{1,2,3}, 
Shigeo S Kimura\altaffilmark{2,4},
Zolt\'an Haiman\altaffilmark{1}
}
\affil{
\altaffilmark{1}Department of Astronomy, Columbia University, 550 W. 120th St., New York, NY, 10027, USA\\
\altaffilmark{2}Astronomical Institute, Graduate School of Science, Tohoku University, Aoba, Sendai 980-8578, Japan\\
\altaffilmark{3}National Astronomical Observatory of Japan, National Institutes of Natural Sciences, 2-21-1 Osawa, Mitaka, Tokyo 181-8588, Japan\\
\altaffilmark{4}Frontier Research Institute for Interdisciplinary Sciences, Tohoku University, Sendai 980-8578, Japan\\
}
\email{E-mail: hiromichi.tagawa@nao.ac.jp}

\begin{abstract} 
Some Seyfert galaxies are detected in high-energy gamma rays, but the mechanism and site of gamma-ray emission are unknown. Also, 
the origins of the cosmic 
high-energy neutrino and MeV gamma-ray 
backgrounds have been veiled in mystery since their discoveries. 
We propose 
emission from stellar-mass BHs (sBHs) embedded in disks of active galactic nuclei (AGN) as their possible sources. 
These sBHs are predicted to launch jets due to the Blandford-Znajek mechanism, which can produce intense electromagnetic, neutrino, and cosmic-ray emissions. 
We investigate whether these emissions can be the sources of cosmic high-energy particles. 
We find that emission from internal shocks in the jets 
can explain gamma rays from nearby radio-quiet Seyfert galaxies including NGC1068, 
if the Lorentz factor of the jets ($\Gamma_{\rm j}$) is high. 
On the other hand, for moderate $\Gamma_{\rm j}$, 
the emission 
can significantly 
contribute to the background gamma-ray and neutrino intensities in the $\sim {\rm MeV}$ and $\lesssim {\rm PeV}$ bands, respectively. 
Furthermore, for moderate $\Gamma_{\rm j}$ with efficient amplification of the magnetic field 
and cosmic-ray acceleration, 
the neutrino emission from NGC1068 and the ultrahigh-energy cosmic rays can be explained. 
These results suggest that the neutrino flux from NGC1068 as well as the background intensities of ${\rm MeV}$ gamma rays, neutrinos, and the ultrahigh-energy cosmic rays can be explained by a unified model. 
Future MeV gamma-ray satellites will test our scenario for neutrino emission. 
\end{abstract}

\section{Introduction}

Our Universe is filled with high-energy particles, including charged particles (cosmic rays; CRs), neutrinos, and gamma rays, but the origins of these cosmic high-energy particles are unknown.
Recently, significant progress 
have been made in high-energy neutrino astrophysics.
IceCube reported the detection of extraterrestrial neutrinos in 2013 \citep{IceCube2013diffuse}, and has been improving the measurement of the cosmic high-energy neutrino background for 10 TeV--100 PeV \citep{Aartsen2015,Aartsen2020}. 
They also identified a nearby Seyfert galaxy, NGC 1068, as a cosmic neutrino source \citep{Aartsen2020_Neutrino_NGC1068,IceCube2022_NGC1068}. In addition, they reported a hint of association between neutrino signals and radio-quiet active galactic nuclei (AGN) \citep{Abbasi2021_AGN_Neutrino}. 
However, possible mechanisms for CR acceleration and subsequent neutrino production sites in radio quiet AGN remain unclear. This has motivated investigations of non-thermal phenomena in hot accretion flows \citep{Kimura2021_llagn,Gutierrez2021}, hot coronae \citep{Murase2020,Eichmann22}, accretion shocks \citep{Inoue2020}, disk winds \citep{InoueS22_Wind}, and jets from accreting binaries \citep{Sridhar2022}.

High-energy gamma rays ($>100$ MeV) have also been detected from radio-quiet AGNs \citep{Wojaczynski2015}, but the origin of these gamma-rays are also controversial. The majority of these gamma-ray detected AGNs exhibit signatures of starburst activity, which causes gamma-ray production via hadronuclear interactions \citep[e.g.,][]{Ajello2020starburst}. The neutrino-emitting AGN, NGC 1068, is also detected in high-energy gamma rays and shows starburst activity. 
On the other hand, the gamma-ray signals from NGC 4945 show spectral variation correlated with its X-ray flux, which implies that gamma rays may be associated with AGN activity \citep{Wojaczynski2017}. AGN wind interacting with the dusty torus can be a possible site of high-energy gamma-ray production \citep{InoueS22_Wind}. 

Finally, the origins of the unresolved cosmic MeV gamma-ray background and ultrahigh-energy cosmic rays (UHECRs) have been unknown for a long time \citep{Inoue2014,AlvesBatista2019}. Radio-quiet AGNs have been proposed as  candidates for both the cosmic MeV gamma-ray background \citep[e.g.,][]{Inoue2013,Kimura2021_llagn} and a source of UHECRs \citep{Peer2009}.  

Here, we propose a novel scenario for high-energy emission from radio-quiet AGNs, where we consider relativistic jets launched from stellar-mass black holes (sBHs) embedded in AGN disks. 
It has been predicted that stars and compact objects (COs) including stellar-mass black holes (sBHs) are embedded in AGN disks due to capture by dynamical interactions of nuclear star clusters (\citealt{MiraldaEscude00,Lu13}) with the AGN disk \citep{Ostriker1983,Syer1991} and in-situ star formation \citep{Levin2003,Goodman04,Thompson05,Levin2007}. There are several observations supporting this picture \citep{Artymowicz1993,Levin2003,Milosavljevic2004,Tagawa19}. 
Recently, the evolution of COs in AGN disks 
has attracted significant 
attention as these are promising environments for some of the sBH-sBH  \citep[e.g.][]{Bartos17,Stone17,McKernan17,Yang19b_PRL,Tagawa19} 
and sBH-neutron star (NS) mergers \citep{McKernan2020_BHNSWD,Tagawa20_MassGap} reported as gravitational wave (GW) events by LIGO/Virgo/KAGRA \citep{LIGO20_O3_Catalog,Abbott21_GWTC3}.

Many recent studies in the wake of the LIGO/Virgo/KAGRA discoveries have investigated emission from transients emerging in AGN disks \citep{McKernan2019_EM,Graham20,Perna2021_GRBs,Perna2021_AICs,Zhu2021_Cocoon_NSMs,Zhu2021_Neutrino,Zhu2021_WD_AIC,Yang2021_TDE,Moranchel-Basurto2021,Grishin2021,Kimura2021_BubblesBHMs,Wang2021b,Yuan2021}. 
Closest to the present study,
\citet{Wang2021_TZW} considered emission from shocks emerging from interactions of Blandford-Znajek (BZ) jets, \citep{Blandford1977} launched from accreting sBHs, with gas in the AGN's broad line region. In \citet{Tagawa2022_BHFeedback} (Paper~I), we estimated the structure of the cavity created by the BZ jet and the dynamical evolution of gas around the BHs.  In
\citet{Tagawa2023} (Paper~II), we investigated the properties of emission released when a jet, launched from merging sBHs embedded in an AGN disk, breaks out from the disk. 
In this paper, 
we consider in more detail the high-energy radiation from jets launched from solitary sBHs in AGN disks. Non-thermal electrons accelerated at the internal shock emit broadband electromagnetic radiation via synchrotron and inverse Compton scattering. Non-thermal protons, accelerated together with electrons, will produce neutrinos via hadronic interactions and might become cosmic rays after escaping from the system.
We evaluate the possibility that gamma rays and neutrinos from nearby Seyfert galaxies, including NGC1068, are produced in such jets. 
We also estimate their contributions to the 
diffuse cosmic gamma-ray, neutrino, and UHECR background intensities.

\begin{figure}
\begin{center}
\includegraphics[width=65mm]{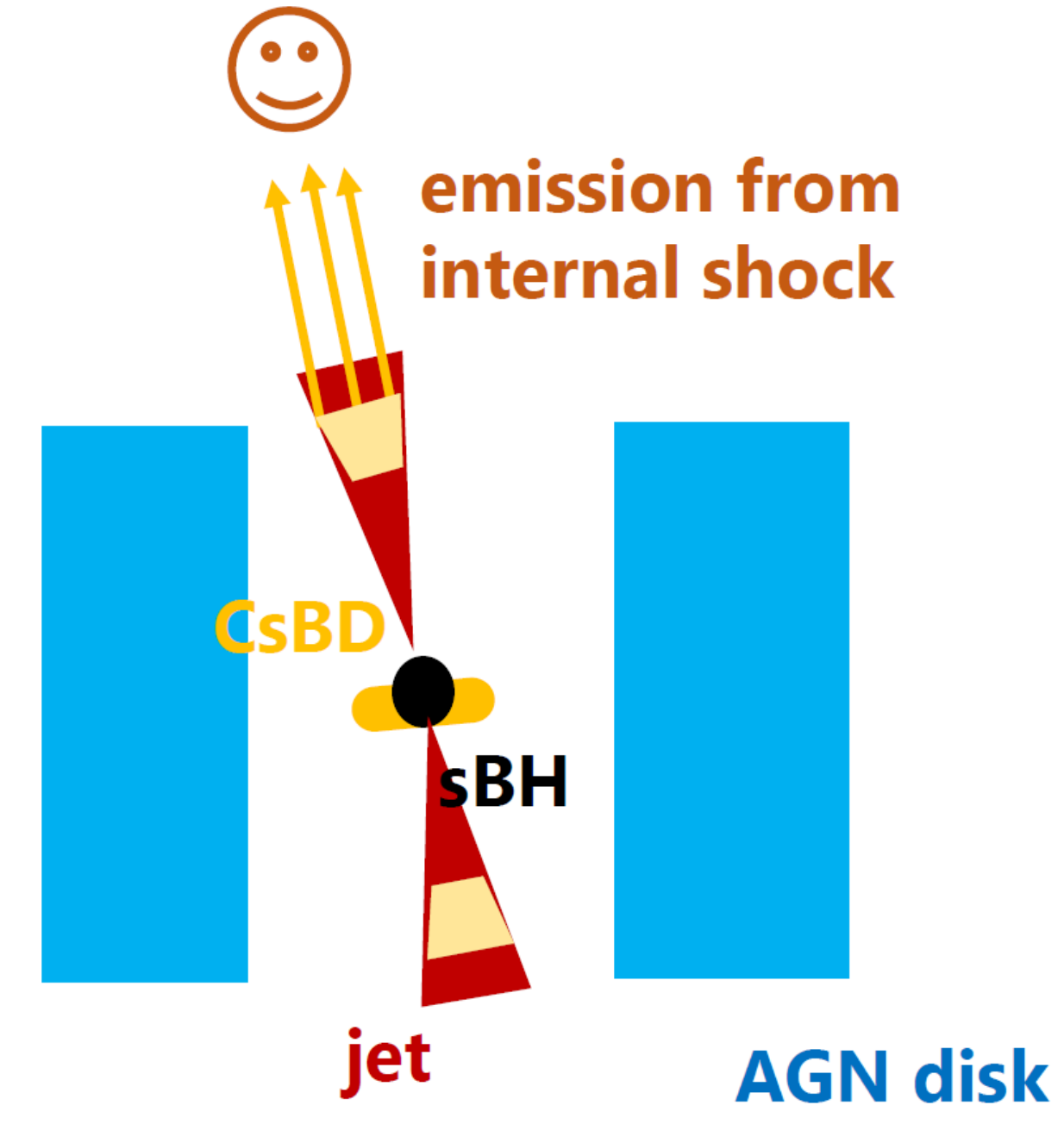}
\caption{
A schematic picture of emission from internal shocks in a jet launched from an sBH accreting gas in an AGN disk. 
}
\label{fig:schematic}
\end{center}
\end{figure}

\section{Model}
\label{sec:model}

To assess the observability of electromagnetic (EM), neutrino, and CR emission produced in shocks around 
the BZ jets, 
we first model the properties of the jets launched from rapidly accreting and spinning sBHs in AGN disks (see Fig.~\ref{fig:schematic} for a schematic illustration). 
The jet is predicted to be launched from an accreting sBH embedded in an AGN disk (Appendix~\ref{sec:accretion_remnants}). 
We focus on sBHs, which are expected to be common in AGN disks and presumably dominate the total jet luminosity,  and
we adopt the same values for our model parameters as in the fiducial model of Paper~I. 
For these parameters, 
through Eq.~\eqref{eq:l_j} and Eq.~(1) of Paper~I, the jet power of $L_{\rm j}\sim 2\times 10^{42}~{\rm erg/s}$ is derived, which we adopt as a fiducial value.

In the fiducial model (model~M1), 
we set 
the fraction of 
postshock energy 
carried by the post-shock magnetic field and by electrons and protons to 
$\epsilon_{\rm B}=0.01$, 
$\epsilon_{e}=0.02$, 
$\epsilon_{\rm CR}=0.05$ (e.g., \citealt[][]{Panaitescu2001,Santana2014,Spitkovsky2008b,Sironi_2013,Tomita2019}), 
respectively; 
the power-law slope for injected electrons and protons accelerated by the first order Fermi process to $p=2.5$;
the Lorentz factor of the jet to $\Gamma_{\rm j}=30$ as derived in Appendix~\ref{sec:properties_jets};
the variability timescale of the jet to $T_{\rm vari}=10^{-3}~{\rm s}$;
and 
the opening angle of the injected jet to $\theta_{\rm j}=0.2$ \citep[e.g.][]{Pushkarev2009,Hada2013,Hada2018,Berger2014}. 
Here note that the parameters $\epsilon_{\rm B}$, $\epsilon_{e}$, 
$\epsilon_{\rm CR}$, 
$p$, and $\Gamma_{\rm j}$ 
are highly uncertain and 
expected to be 
distributed in wide ranges of values as $\epsilon_{\rm B}\sim 10^{-5}$--$0.3$, $\epsilon_{e}\sim 10^{-2}$--$0.5$, 
$\epsilon_{\rm CR}\sim 0.05$--$0.2$, 
$p\sim 2$--$3$, and 
$\Gamma_{\rm j}\sim 2$--$100$ 
depending on sources 
\citep[e.g.][]{Panaitescu2001,Sironi_2013,Santana2014,Caprioli2014,Troja2019,Matsumoto2020,Caprioli2020}. 
We also show the results with $\Gamma_{\rm j}=4$, $\epsilon_{e}=0.05$, $\epsilon_{\rm B}=0.005$, 
$\epsilon_{\rm CR}= 0.05$, 
and $p=2.2$ (model~M2), 
and those with 
$\Gamma_{\rm j}=4$, $\epsilon_{e}=0.05$, $\epsilon_{\rm B}=0.3$, 
$\epsilon_{\rm CR}= 0.15$, $p=2.2$, 
and $L_{\rm j}=10^{43}~{\rm erg/s}$ 
(model~M3), 
as results are sensitive to these parameters 
(see Appendix~\ref{sec:parameter_space} for the parameter space where our non-thermal emission models are applicable. )
Note that $L_{\rm j}=2\times 10^{42}~{\rm erg/s}$ and $L_{\rm j}=10^{43}~{\rm erg/s}$ adopted in models~M1--M2 and M3 are, respectively, predicted for the jets from sBHs at the distance from the SMBH being $1$ and $0.01~{\rm pc}$ (Paper~I), where sBHs are typically accumulated due to a long migration timescale and gap formation \citep{Tagawa19,Perna2021_AICs}. 

\begin{figure}
\begin{center}
\includegraphics[width=85mm]{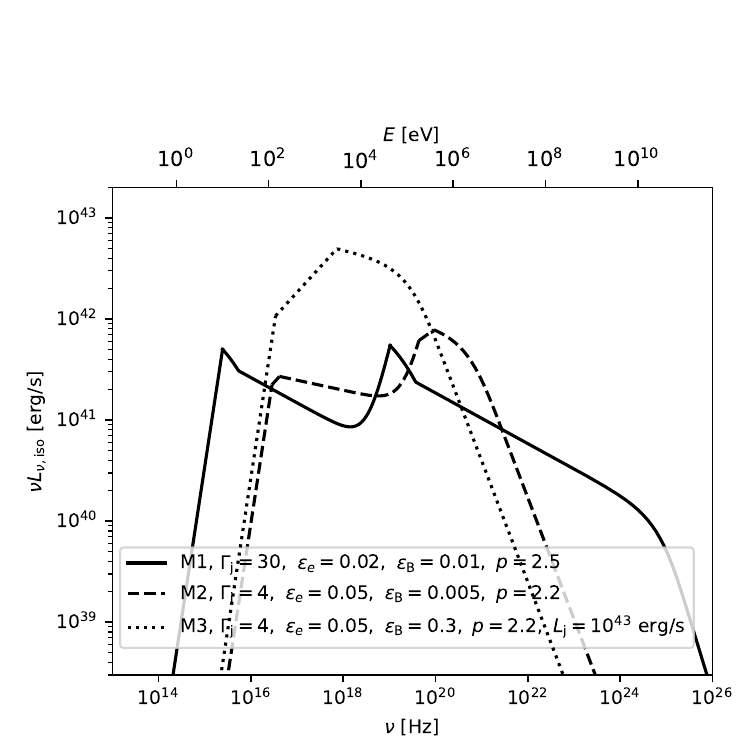}
\caption{
The spectral energy distribution for non-thermal emission from internal shocks of the jet for models~M1 (solid), M2 (dashed), and M3 (dotted). 
}
\label{fig:l_j1}
\end{center}
\end{figure}

\begin{figure}
\begin{center}
\includegraphics[width=85mm]{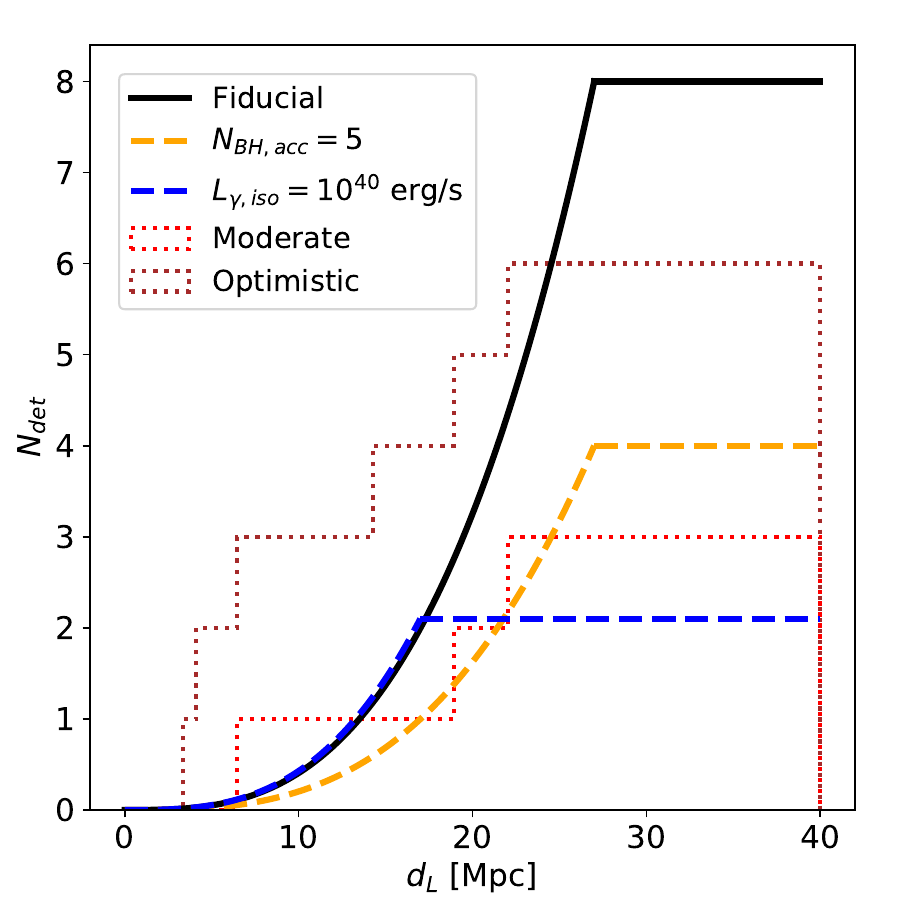}
\caption{
The cumulative number of gamma-ray emitting sources predicted to be detectable by {\it Fermi}-LAT as a function of 
the luminosity distance 
for the fiducial model (solid black line), 
a model with a smaller number of accreting sBHs per AGN disk ($N_{\rm BH,acc}=5$, dashed orange line), 
and that with a weaker gamma-ray luminosity ($L_{\gamma, \rm iso}=10^{40}\,{\rm erg/s}$, dashed blue line). 
The number of gamma-ray sources, inferred from observations, in nearby radio-quiet Seyfert galaxies for moderate and optimistic cases are shown in dotted red and brown lines. 
}
\label{fig:ndet_nbh}
\end{center}
\end{figure}

\section{Gamma rays}

During the propagation of the jet, its kinetic energy is considered to be dissipated. 
We assume that the fraction, $\epsilon_{\rm CR}$ and $\epsilon_{e}$, 
of the postshock energy 
of the jet is used to accelerate protons and electrons, respectively, via the first-order Fermi process 
\citep{Bell1978,Blandford1987_CR}. 
Then, the non-thermal electrons emit broadband radiation via synchrotron and inverse Compton scattering. 
The spectral shapes of the non-thermal emission produced at internal shocks of the jet are presented in Fig.~\ref{fig:l_j1} (see Appendix~\ref{sec:nonthermal} for their derivation). 
In the fiducial model (M1), 
synchrotron emission and synchrotron-self Compton scattering, respectively, produce the bright emission in optical--$\rm MeV$ and in X-ray--$\rm GeV$ bands. 
In addition, synchrotron self-absorption 
and $\gamma\gamma$ annihilation create upper and lower cutoffs, 
respectively. 
The emission by hadronic processes is also significantly absorbed by $\gamma\gamma$ annihilation as discussed 
in Section \S~\ref{sec:hadronic_emission}. 
Below we discuss whether gamma-ray emission from the jets can be detected from nearby Seyfert galaxies. 
In addition, we discuss the contribution of the jets to the diffuse gamma-ray background.

\subsection{Emission from radio-quiet Seyfert galaxies}

We consider the possibility that the gamma rays from radio-quiet galaxies are originated from the BZ jets launched from sBHs in AGN disks. 
According to Fig.~\ref{fig:l_j1}, 
the isotropic-equivalent gamma-ray luminosity from the jet at $\sim 1$--$10~{\rm GeV}$ \citep{Wojaczynski2015} for model~M1 is $L_{\rm GeV, \rm iso}\sim 3\times 10^{40}~{\rm erg~s^{-1}}$.
Then, 
the intrinsic gamma-ray luminosity is 
\begin{eqnarray}
\label{eq:l_gamma_iso}
L_{\rm GeV}= p_\theta L_{\rm GeV,iso}= 
\nonumber\\
\sim 5\times 10^{38}~{\rm erg~s^{-1}}
\left(\frac{L_{\rm GeV,iso}}{3\times 10^{40}~{\rm erg~s^{-1}}}\right)
\left(\frac{\theta_{\rm j}}{0.2}\right)^{2},
\end{eqnarray}
where 
$p_\theta = \theta_{\rm j}^2/2$ is the probability that the jet 
is directed towards an observer. 
By using {\it Fermi}-LAT with the sensitivity of $(E_\gamma F_{E_\gamma})_{\rm LAT}\sim 3\times 10^{-13}~{\rm erg~cm^{-2}~s^{-1}}$ at $\sim {\rm GeV}$ \citep{Atwood2009,Funk2013}, 
the detectable distance for the gamma rays is 
\begin{eqnarray}
\label{eq:dl_det}
d_{L,\rm det}=
\left(\frac{L_{\rm GeV, \rm iso}}{4\pi (E_\gamma F_{E_\gamma})_{\rm sens}}\right)^{1/2} \nonumber\\
\sim 27\,{\rm Mpc}
\left(\frac{L_{\rm GeV, \rm iso}}{3\times 10^{40}\,{\rm erg/s}}\right)^{1/2}
\left(\frac{(E_\gamma F_{E_\gamma})_{\rm LAT}}{3\times 10^{-13}~{\rm erg~cm^{-2}s^{-1}}}\right)^{-1/2}.
\end{eqnarray}
The detectable number of such gamma-ray sources 
within the distance $d_L$ is roughly estimated as 
\begin{eqnarray}
\label{eq:n_det_gamma}
N_{\rm det}(d_L)=
p_{\theta}
\frac{4\pi [{\rm min}(d_L,d_{L,\rm det})]^3 }{3} 
n_{\rm rotBH,acc} ~
\nonumber\\
\sim 8
\left(\frac{\theta_{\rm j}}{0.2}\right)^2
\left(\frac{{\rm min}(d_L,d_{L,\rm det})}{27\,{\rm Mpc}}\right)^3
\left(\frac{n_{\rm rotBH,acc}}{5\times 10^{-3}\,{\rm Mpc}^{-3}}\right) ~~~
\end{eqnarray}
where 
$n_{\rm rotBH,acc}=n_{\rm AGN}N_{\rm rotBH,acc}$ 
is the number density of rapidly accreting and spinning sBHs in AGN disks, 
$N_{\rm rotBH,acc}$ is 
their typical number 
in a single AGN disk
and $n_{\rm AGN}$ is the AGN 
space density. 
In Eq.~\eqref{eq:n_det_gamma}, we assume that the jets (and the sBH spins) are randomly oriented \citep{Tagawa19,Tagawa20b_spin}, and we adopt $n_{\rm AGN}=5\times 10^{-4}\,{\rm Mpc^{-3}}$ considering AGNs with X-ray luminosity of $L_X \gtrsim 10^{42}\,{\rm erg/s}$ \citep{Ueda2014}. 
Assuming
that at any given time, the active fraction of sBHs is $f_{\rm active}\sim 0.1$ as roughly estimated in Paper~I, we adopt
\begin{eqnarray}
\label{eq:N_rotbh_acc}
N_{\rm rotBH,acc}=N_{\rm rotBH,AGN} f_{\rm active} \nonumber\\
\sim 10\left(\frac{N_{\rm rotBH,AGN} }{100}\right)
\left(\frac{f_{\rm active}}{0.1}\right), 
\end{eqnarray}
where $N_{\rm rotBH,AGN}\sim 100$ is based on typical numbers found in models for the AGN-embedded sBH population \citep{Tagawa19,Tagawa20_MassGap}. 
Although the distribution of the positions of sBHs in an AGN disk is uncertain 
and is influenced by phases and properties of AGNs, 
we assume that most sBHs are at $1~{\rm pc}$ (models~M1 and M2) and $\sim 10\%$ of them are at $0.01~{\rm pc}$ (model~M3). 
Here, the former component corresponds to the sBHs captured by an AGN disk before considerable migration, while the latter corresponds to the sBHs accumulated at gaps, which are predicted to form in low opacity regions of an AGN disk \citep{Thompson05,Tagawa19}. 
The low number density of active sBHs at $\sim 0.01~{\rm pc}$ is required so that the neutrino flux for model~M3 does not exceed the background neutrino flux (\S~\ref{sec:neutrino}), 
while if it is lowered further, the probability to detect neutrinos from NGC1068 by model~M3 becomes low (\S~\ref{sec:neutrino_ngc1068}).

Fig.~\ref{fig:ndet_nbh} compares the detectable number of gamma-ray sources predicted by the models to the number of detected gamma-ray sources from nearby radio-quiet Seyfert galaxies observed by {\it Fermi}-LAT. 
As fiducial candidates (dotted red line), 
we include NGC4151, NGC6814, and NGC4258, 
for 
which the likelihood ratio of the non-detection to the detection of gamma rays are $\sim 2\times 10^{-4}$, $\sim 10^{-7}$, and $\sim 9\times 10^{-3}$, respectively. 
In the "optimistic" distribution (dotted brown), 
we additionally include three Seyfert galaxies 
with 
starbursts, Circinus, NGC1068, and NGC4945, 
as the gamma rays from these galaxies are not fully explained by 
starbursts \citep{Hayashida2013,Eichmann2016,Wojaczynski2017}. 
From the figure, we find that the observed 
distribution of gamma-ray sources
is consistent with our models 
with
$n_{\rm rotBH,acc}\sim 2\times 10^{-3}$ -- $5\times 10^{-3}~{\rm Mpc}^{-3}$ 
and $L_{\rm GeV, \rm iso}\sim 10^{40}$ -- $3\times 10^{40}\,{\rm erg/s}$ (solid black, dashed blue, and dashed orange lines). 
Conversely, 
if the gamma rays from the radio-quiet Seyfert galaxies are not originated from sBHs (e.g., $N_{\rm det}\sim 0$),  
$n_{\rm rotBH,acc}$ can be roughly constrained to be 
\begin{eqnarray}
\label{eq:n_rot_inv}
n_{\rm rotBH,acc}
\lesssim 
5\times 10^{-4}\,{\rm Mpc}^{-3}
\left(\frac{N_{\rm det}}{1}\right)
\left(\frac{\theta_{\rm j}}{0.2}\right)^{-2}\nonumber\\
\left(\frac{L_{\rm GeV,iso}}{3\times 10^{40}~{\rm erg/s}}\right)^{-3/2}
\left(\frac{(E_\gamma F_{E_\gamma})_{\rm LAT}}{3\times 10^{-13}~{\rm erg~cm^{-2}s^{-1}}}\right)^{3/2}.
\end{eqnarray}
Note that the gamma-ray luminosity at $\sim$ GeV bands is obscured by $\gamma \gamma$ annihilation for low Lorentz factors (e.g. dashed and dotted lines with $\Gamma_{\rm j}=4$ in Fig.~\ref{fig:l_j1}).
In this case, the jets from sBHs in AGN disks cannot explain the GeV gamma rays from Seyfert galaxies, and another explanation would be required.

\begin{figure}
\begin{center}
\includegraphics[width=85mm]{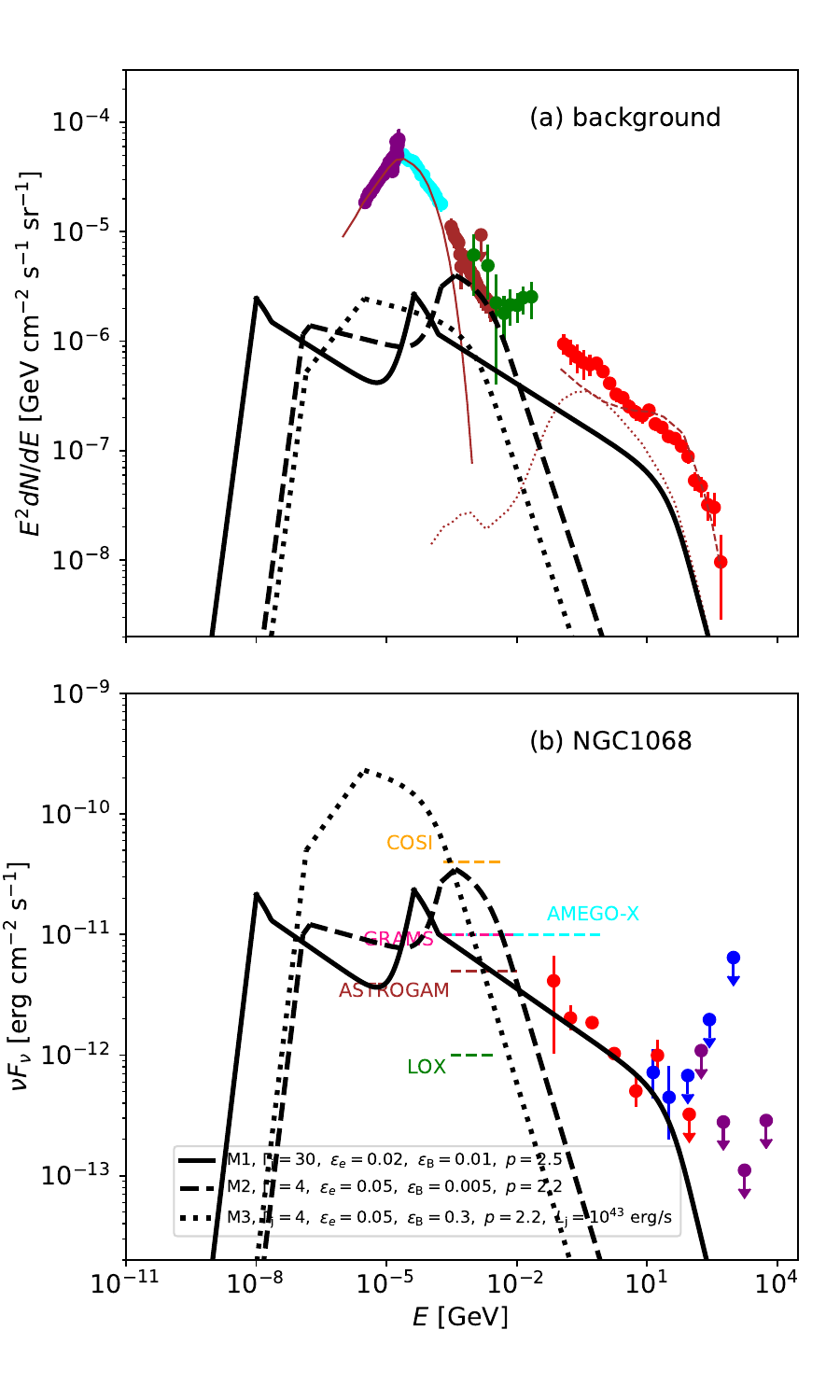}
\caption{
The contribution to the gamma-ray flux from NGC1068 (panel~b) and the background gamma-ray intensity (panel~a) by the internal shocks 
for 
models~M1 (solid black), 
M2 (dashed black), 
and M3 (dotted black). 
In panel~a, 
solid points represent the intensity observed 
by Fermi-LAT (red, \citealt{Ackermann2015_diffuse}), 
COMPTEL (teal, \citealt{Weidenspointner2000}), 
SMM (brown, \citealt{Watanabe1997}), 
Swift BAT (cyan, \citealt{Ajello2008}),
and RXTE (purple, \citealt{Revnivtsev2003}). 
Thin solid, dashed, and dotted brown lines 
are the intensities predicted by models for 
Seyfert galaxies \citep{Gilli2007}, blazars \citep{Giommi2015}, and star-forming galaxies \citep{Lacki2014}, presented in \citet{DeAngelis2021}. 
In panel~b, 
the red, blue, and purple points
are the observed gamma-ray intensities adopted from \citet{Abdollahi2020}, \citet{Ajello2017}, and \citep{Acciari2019}, respectively. 
The sensitivities for future telescopes in $\sim {\rm MeV}$ bands are presented by colored dashed lines. 
}
\label{fig:gamma_intensity}
\end{center}
\end{figure}

\subsection{Cosmic gamma-ray backrground intensity}

We next estimate 
the contribution of gamma-ray emission from 
the global population of 
jets to the diffuse background intensity in MeV bands as 
\begin{eqnarray}
\label{eq:e_gamma}
E_\gamma^2 \Phi_\gamma 
\sim  f_{z} 
L_{\rm MeV,iso}p_\theta 
n_{\rm rotBH,acc}\frac{c}{4\pi H_0}
\nonumber\\
\sim 5\times 10^{-6} {\rm GeV~cm^{-2}~s^{-1}~sr^{-1}}\nonumber\\ \nonumber \hspace{0.5\baselineskip}\\
\times \left(\frac{L_{\rm MeV,iso}p_\theta}{2\times 10^{40}\,{\rm erg/s}}\right)
\left(\frac{n_{\rm rotBH,acc}}{5\times 10^{-3}\,{\rm Mpc^{-3}}}\right)
\left(\frac{f_{z}}{2}\right) 
\end{eqnarray}
(Fig.~\ref{fig:gamma_intensity}a), 
where $L_{\rm MeV,iso}$ is the isotropic-equivalent gamma-ray luminosity around MeV bands,   $H_0=67.8\,{\rm km~s^{-1}~Mpc^{-1}}$ is the Hubble constant \citep{Planck2016} , 
and  $f_z = 2$ is a correction factor for redshift evolution (Appendix~\ref{sec:redshift}). 
We find that the gamma-ray flux 
in our fiducial model is generally an order of magnitude or more below the observed background intensity.  However, model~M2 can explain the gamma-ray background intensity in the narrow $\sim 1$--$10~{\rm MeV}$ bands (dashed black line and brown points in Fig.~\ref{fig:gamma_intensity}a, \citealt{Ackermann2015_diffuse}). The origin of the background in this energy range has not been understood, as other, 
previously proposed astrophysical contributions significantly underpredict the level of the background (see thin solid, dashed, and dotted brown lines in Fig.~\ref{fig:gamma_intensity}a).
It is notable that our model can also explain the neutrino background intensities as shown 
in Section $\S~\ref{sec:UHECRs_Neutrinos}$ below.

\subsection{Emission from NGC1068}

We next consider whether the emission from internal shocks of the jets can explain the gamma-ray emission from NGC1068. 
Note that 
NGC1068 is a type~II AGN, and 
its intrinsic X-ray luminosity 
is $\sim 7\times 10^{43}~{\rm erg/s}$ \citep{Marinucci2016}, 
which is significantly brighter than the X-ray emission by the jets from sBHs in the AGN disk (Fig.~\ref{fig:l_j1}).

Fig.~\ref{fig:gamma_intensity}b compares the observed gamma-ray flux from NGC1068 and the predicted fluxes in 
models~M1--M3. 
Model~M1 (solid line) significantly contributes to the gamma-ray emission from NGC1068 
between $\sim$100 MeV to $\sim$100 GeV 
because model~M1 avoids $\gamma\gamma$ annihilation owing to the high Lorentz factor.
On the other hand, 
models~M2 and M3 have 
not been constrained by the current gamma-ray observations. 
Also, the emission in infrared to X-ray bands is significantly absorbed by dust (although this effect is not incorporated in the predictions in Fig.~\ref{fig:gamma_intensity}~b) and the fluxes in the $\sim {\rm keV}$ -- $\rm MeV$ bands receive significant contributions from coronae in the AGN. 
Thus, 
to test 
models~M2 and M3, 
MeV gamma rays are useful. 
MeV gamma-rays can be detected with future gamma-ray telescopes, such as 
the Compton Spectrometer and Imager (COSI) \citep{Tomsick2019_COSI}, 
the All-sky Medium Energy Gamma-ray Observatory eXplorer (AMEGO)\citep{Caputo2022_AMEGO}, 
Gamma-Ray and AntiMatter Survey (GRAMS) \citep{Aramaki2020_GRAMS}, 
eASTROGAM\citep{DeAngelis2021}
and 
the Lunar Occultation eXplorer (LOX)\citep{Miller2019_LOX}.

\begin{figure}
\begin{center}
\includegraphics[width=85mm]{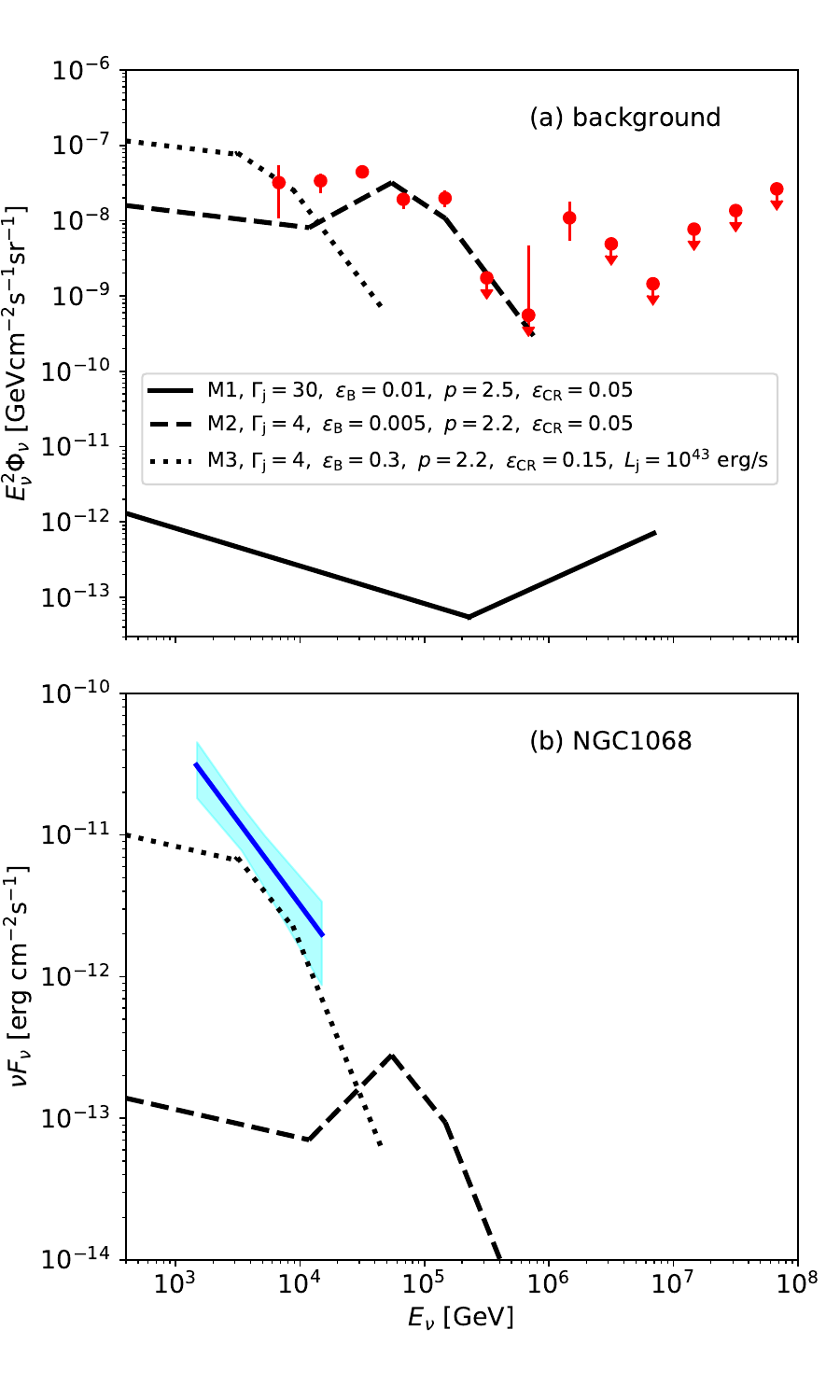}
\caption{
The contribution to the the background intensity (panel~a) and the flux from NGC1068 (panel~b) 
for a single neutrino flavor 
by internal shocks for models~M1 (solid black), M2 (dashed black), and M3 (dotted black).  
The observed neutrino intensity \citep{Aartsen2020} is presented by red points (panel~a), and 
the blue shaded region represents the 1, 2, and 3$~\sigma$ uncertainty on the spectrum measured by \citet{Aartsen2020_Neutrino_NGC1068} (panel~b). 
}
\label{fig:ephi_j1}
\end{center}
\end{figure}

\section{UHECRs and Neutrinos}

\label{sec:UHECRs_Neutrinos}

In this section, we estimate whether high-energy protons and neutrinos produced from the jets can explain the observed background fluxes and the neutrino flux from NGC1068.

\subsection{UHECRs}

Since the adiabatic expansion is the most efficient cooling process for protons accelerated in the jets in the models, 
the maximum proton energy accelerated in internal shocks of the jets is 
given by the comparison between the acceleration timescale in the Bohm limit (in which the particle mean free path is assumed to be equal to the Larmor radius) and the expansion timescale as 
\begin{eqnarray}
\label{eq:e_max_p}
E_{p,\rm max}=E'_{p,\rm max}\Gamma_j =e B_{\rm j}' Z_{\rm diss}
 \nonumber\\
 \sim 10^{19} {\rm eV} 
\left(\frac{B_{\rm j}' }{4\times 10^7~{\rm G}}\right)
\left(\frac{Z_{\rm diss} }{1\times 10^{9}~{\rm cm}}\right)
\end{eqnarray}
where $B_{\rm j}'$ is the magnetic field strength of the shocked jet, $Z_{\rm diss}=2 T_{\rm vari} c \Gamma_{\rm j}^2$ is the typical dissipation radius for internal shocks, $e$ is the electric charge, 
and the primes denote quantities in the fluid comoving frame. 
The maximum proton energies for 
models~M1, M2, and M3 are, respectively, $1\times 10^{17}$, $7\times 10^{17}$ and 
$1\times 10^{19}~{\rm eV}$. 
It is notable that the CR production at a few $10^{17}~{\rm eV}$ predicted in model~M1 may be related to the observations of CRs around this energy band \citep{Buitink2016}. 
On the other hand, to produce high $E_{p, \rm max}$, low $\Gamma_{\rm j}$ and high $\epsilon_{\rm B}$ are required, and model~M3 can explain the energy of the UHECRs.

The background intensity of UHECRs produced in the jets from sBHs in AGN disks is estimated as   
\begin{eqnarray}
\label{eq:e_diff_cr}
E_p ^2 \Phi_p
\sim 
\frac{\epsilon_{\rm CR} L_{\rm j}}{R_p}
n_{\rm rotBH,acc}f_{z} \frac{c }{4\pi H_0}
 \nonumber\\
\sim 1\times 10^{-7} {\rm GeV~cm^{-2}~s^{-1}~sr^{-1}} \nonumber\\ \vspace{0.5\baselineskip} \nonumber \\
\times\left(\frac{\epsilon_{\rm CR} }{0.15}\right)
\left(\frac{L_{\rm j}}{10^{43}\,{\rm erg/s}}\right)\nonumber\\
\times\left(\frac{n_{\rm rotBH,acc}}{5\times 10^{-4}\,{\rm Mpc^{-3}}}\right)
\left(\frac{R_p}{300}\right)^{-1}
\left(\frac{f_z}{2}\right), 
\end{eqnarray}
where $E_p$ is proton energy, 
$R_p^{-1}=(E_p/E_{p,\rm min})^{2-p}(p-2)/[1-(E_{p,\rm min}/E_{p,\rm max})^{p-2}]$, 
and $E_{p,\rm min}=\Gamma_{\rm j} m_p c^2$ 
is the minimum proton energy. 
This value is consistent with the observed value for the cosmic-ray intensity of $\sim 10^{-7}~{\rm GeV~cm^{-2}~sr^{-1}~s^{-1}}$ at $E_p \sim 10^{18}~{\rm eV}$ \citep{Blasi2013,Zweibel2013}, while it is not obvious whether extragalactic CRs of $E_p<10^{18}$ eV can propagate to the Earth within the Hubble time \citep{Kimura2015}. 
Since the observed cosmic-ray background intensity of $\lesssim 10^{-8}~{\rm GeV~cm^{-2}~sr^{-1}~s^{-1}}$ at $E_p \sim 10^{20}~{\rm eV}$ \citep[e.g.][]{Zweibel2013} is lower 
by about an order of magnitude 
compared to the production rate in models~M1--M3, and the energy can be predicted by model~M3, the UHECRs could be explained by model~M3. 
Also, the high density of sBHs in AGN disks 
($\sim 5\times 10^{-4}$--$5\times 10^{-3}~{\rm Mpc}^{-3}$) 
is consistent with the source density expected for producing the UHECRs \citep{Takami2009}. 
On the other hand, 
since the heavy metal abundance ratio of the UHECR compositions \citep{Aab2014} is higher than that of AGNs \citep{Shields1996,Xu18}, 
we need to consider some heavy-element enhancement process, such as re-acceleration of galactic CRs \citep{Caprioli2015,Kimura2018_UHECR,Zirakashvili2022}. 
Since the cosmic-ray flux and abundance ratio are expected to be changed during propagating to the Earth, 
detailed consistency of the flux and the metal abundance distribution expected in this model are worth investigating in the future.

\subsection{Cosmic neutrino background}

\label{sec:neutrino}

Interactions of high-energy protons lead to the production of pions, which generate neutrinos via decay processes. 
We discuss the contribution of such neutrino emission to the background neutrino flux and the neutrino flux from NGC1068.

As cooling processes for baryons, we consider adiabatic cooling, $pp$ reaction, and $p\gamma$ reaction, 
where pions could be produced by inelastic nuclear $pp$ reactions and photohadronic ($p\gamma$) reactions, respectively, 
and the pions subsequently decay into neutrinos. 
Using the fraction of protons producing pions through $pp$ and $p\gamma$ reactions ($f_{pp}$ and $f_{p\gamma}$, Appendix~\ref{sec:neutrino_production}), the jet power, 
and the suppression factor due to pion cooling $f_{\pi, \rm sup}$ \citep{Kimura2022_NeutrinoGRB}, the diffuse neutrino flux produced from internal shocks of jets can be roughly estimated as (e.g., \citealt{Razzaque2004,Murase2016})
\begin{eqnarray}
\label{eq:e_diff_neutrino}
\epsilon_\nu ^2 \Phi_\nu
\sim \frac{3K}{4(1+K)} \frac{{\rm min}[1,{\max}(f_{p\gamma},f_{pp})]f_{\pi, \rm sup}
\epsilon_{\rm CR} L_{\rm j}}{
R_p
}\nonumber\\
n_{\rm rotBH,acc}f_{z} \frac{c }{4\pi H_0}
 \nonumber\\
\sim 1\times 10^{-13} {\rm GeV~cm^{-2}~s^{-1}~sr^{-1}} \nonumber\\
\left(\frac{\epsilon_{\rm CR} }{0.05}\right)
\left(\frac{f_{pp} }{3\times 10^{-6}}\right)
\left(\frac{f_{\pi, \rm sup} }{1}\right)
\left(\frac{L_{\rm j}}{2\times 10^{42}\,{\rm erg/s}}\right)\nonumber\\
\left(\frac{n_{\rm rotBH,acc}}{5\times 10^{-3}\,{\rm Mpc^{-3}}}\right)
\left(\frac{R_p}{500}\right)^{-1}
\left(\frac{f_z}{2}\right)
\end{eqnarray}
(solid black lines in Fig.~\ref{fig:ephi_j1}), 
where $K=1$ and $K=2$, respectively, denote the average ratio of charged to neutral pion for photohadronic ($p\gamma$) and inelastic hadronuclear ($pp$) reaction, and $f_{\pi, \rm sup}$ is the fraction of the energy loss before pions decay. 
Since the neutrino background intensity at $\sim 10^5~{\rm GeV}$ is observed to be $\sim 10^{-7}$--$10^{-8}~{\rm GeV~cm^{-2}~s^{-1}~sr^{-1}}$ (red points in Fig.~\ref{fig:ephi_j1}a, \citealt{Aartsen2015,Aartsen2020}), 
neutrinos produced by the internal shocks 
make only a minor contribution 
to the background neutrino intensity in the fiducial model. 
On the other hand, for low $\Gamma_{\rm j}$, 
$f_{pp}$ and $f_{p\gamma}$ are high, and then the internal shocks could significantly contribute to the background intensity (dashed and dotted black lines in Fig.~\ref{fig:ephi_j1}a). 
As AGNs are expected to be major production sites for neutrinos \citep{Bartos2021} especially around $E_\nu\sim 100~{\rm TeV}$ \citep{Abbasi2021_AGN_Neutrino}, the model with the jets from sBHs in AGN disks may be a promising scenario for producing the background neutrinos in the Universe.

\subsection{Neutrino emission from NGC1068}

\label{sec:neutrino_ngc1068}

We here discuss the possibility that the neutrinos from NGC1068 are produced at internal shocks of jets launched from sBHs embedded in AGN disks. 
Fig.~\ref{fig:ephi_j1}b shows the neutrino flux from one jet in our models 
(models~
M2 and M3) 
and the observed neutrino flux from NGC1068. 
The neutrino flux predicted by 
model~M2 
is not high enough to explain the observed flux 
(dashed black line). 
In the model with high $\epsilon_{\rm B}$ (model~M3), pion cooling via synchrotron radiation 
is efficient, 
and therefore the neutrino flux at high energies in $E_\nu \gtrsim 10^4~{\rm GeV}$ are suppressed. On the other hand, the neutrino flux at lower energies in $E_\nu \lesssim 10^4~{\rm GeV}$, where pion cooling is inefficient, is high, which is consistent with the observed neutrino flux from NGC1068. 
Additionally, to reproduce the neutrino flux from NGC1068, 
high $\epsilon_{\rm CR}$ and $L_{\rm j}$ are required. 
In the high energy ranges around $\sim 10^4~{\rm GeV}$, where the atmospheric backgrounds are much smaller, the observed flux is roughly consistent with the prediction by model~M3.

Although different values of $\epsilon_{\rm B}$ 
and $\epsilon_{\rm CR}$ (in addition to $L_{\rm j}$) 
are required to explain the cosmic neutrino background and the neutrino emission from NGC 1068, this is not unreasonable, since we expect a broad distribution  of $\epsilon_{\rm B}$ 
and $\epsilon_{\rm CR}$. 
Multi-wavelength fits of gamma-ray burst afterglows revealed that the distribution of $\epsilon_{\rm B}$ is indeed very broad \citep{Panaitescu2001,Santana2014} 
and $\epsilon_{\rm CR}$ is observationally less constrained.\footnote{Theoretical models for magnetic amplification 
and cosmic-ray acceleration 
in collisionless shocks are still incomplete with the current computational resources \citep[e.g.][]{Caprioli2014,Caprioli2020,Hu2022,Tomita2022}, 
and we cannot predict 
the values of $\epsilon_{\rm B}$ and $\epsilon_{\rm CR}$ 
in realistic astrophysical environments from the first-principle calculations.} 
If the peaks of the $\epsilon_{\rm B}$ and $\epsilon_{\rm CR}$ distributions 
are as low as in model~M2 and the dispersions are large enough to contain a value as high as in model M3, 
the neutrino emission from NC1068, and the background intensities of gamma rays in $\sim {\rm MeV}$ bands, neutrinos in $\lesssim {\rm PeV}$ bands, and UHECRs are all explained by a single unified model. 
This has not been proposed so far, although we need to change the several parameters between models~M2 and M3. 
Also, 
the high value for $\epsilon_{\rm B}$, required to explain the neutrino flux from NGC1068, may be difficult to be achieved. 
A possible 
issue is that the jet produced by an sBH at $\sim 0.01~{\rm pc}$ in NGC1068 needs to be directed to us. 
Since the SMBH mass and the accretion rate of NGC1068 are high \citep{Pier1994,Greenhill1997} and the number of sBHs in an AGN disk is roughly proportional to the square root of the SMBH mass and the accretion rate  \citep{Tagawa19,Tagawa20_MassGap}, 
the number of sBHs in an AGN disk in NGC1068 is higher by a factor of $\sim 14$ compared to that in the fiducial model. 
Then, the probability that jets produced from sBHs at $\sim 0.01~{\rm pc}$ in NGC1068 are directed to us is $\sim 0.3$, 
which is viable. 
On the other hand, this also means that jets from $\sim 3$ sBHs at $\sim 1~{\rm pc}$ are directed to us, 
which may be problematic to 
be consistent with the X-ray flux from NGC1068. 
This is because the X-ray emission from corona in NGC1068 
may be mostly absorbed at sub-parsec scales, instead of several parsec scales, 
by considering the amount of mass estimated in parsec scales \citep{GarcaBurillo2016,Imanishi2018,Imanishi2020}. 
In that case, the X-ray emission from jets at $\sim 1~{\rm pc}$ may not be fully absorbed and significantly exceed the observed X-ray flux. 
Hence, dedicated estimates 
may be required to clarify whether or not this scenario is consistent with hard X-ray observations by NuSTAR \citep{Marinucci2016,Zaino2020}.

\section{Discussions}

\subsection{Fermi bubble}

Two large gamma-ray bubbles, the so-called Fermi bubbles, have been discovered above and below the center of our Galaxy by gamma-ray \citep{Su_2010}, X-ray \citep{Bland-Hawthorn2003}, and radio telescopes \citep{Finkbeiner2004}. 
The size of the bubble is $\sim 10~{\rm kpc}$, its shape is almost mirror-symmetric with respect to the Galactic plane, and its origin is unknown \citep{YangRuszkowski2018}. 
We discuss whether the Fermi bubble may be related to sBHs in an AGN disk. 
One of promising models for the origin of the Fermi bubble is the leptonic jet model, in which 
a jet is estimated to be launched from the central SMBH $1$--$3$ Myr ago with the duration of $\sim 0.1$--$0.5~{\rm Myr}$ \citep{YangRuszkowski2012,Guo_2012,YangRuszkowski2022}, and the power of the jet is roughly estimated to be $\sim 10^{43}$--$10^{44}~{\rm erg/s}$  \citep{Guo_2012,YangRuszkowski2022}. 
We here propose that it is possible to produce a similar bubble by jets launched from sBHs embedded in an AGN disk. 
In our fiducial model, 
the power of the jet launched from an sBH is 
$L_{\rm j}\sim 10^{42}$--$10^{43}~{\rm erg/s}$, 
and the active number of the jets is $N_{\rm rotBH,AGN}\sim 10$ (Eq.~\ref{eq:N_rotbh_acc}). 
Thus, the total power of the jets launched from the sBHs is  
consistent with the power required to explain the properties of the Fermi bubble. 
Here we note that the symmetry of the Fermi bubble is perhaps surprising, given that SMBH jets in general do not point perpendicular to the galaxy disk \citep{Kinney2000,Hopkins2012}. 
In the jets from the sBHs, the shocked gas is predicted to be symmetric with respect to the galactic plane, if a pc-scale disk is aligned to the galactic plane \citep{Tagawa20b_spin}. 
Since the angular momentum direction of the pc-scale disks tend to be aligned to that of the nuclear star clusters due to vector resonant relaxation \citep{Kocsis2011,Impellizzeri2019,Levin2022}, 
the alignment of the pc-scale disk presumably occurs if nuclear star clusters rotate in the same direction with the host galaxies, which is often observed including our Galaxy (\citealt{Levin2003,Yelda14,Do20,Neumayer20}, but see also \citealt{Kormendy2013} for NGC4258 as a counter-example). 
Also, the duration of each jet from an sBH is $\lesssim 10^3~{\rm yr}$ (Paper~I), which may be consistent with the uniform haze emission, 
as \citet{YangRuszkowski2022} suggested that multiple jets may be required to explain the spatial distribution of the microwave emission without being suppressed by magnetic pressure in the vicinity of the Galactic center. 
Thus, jets from sBHs embedded in an AGN disk may be responsible for producing such bubbles. 
Note that due to their intermittence and short duration,
the jets likely pass through high-pressure regions produced by previous jets and could be significantly decelerated before interacting with the inter-stellar medium. In this case, non-thermal emission from external shocks of the jets might be inefficient. Such emission 
merits investigating 
in the future.

\subsection{Hadronic gamma-ray emission}

\label{sec:hadronic_emission}

When hadronuclear and photohadronic processes produce high-energy neutrinos in astrophysical environments, these processes inevitably produce gamma rays whose energy and luminosity are comparable to those of neutrinos. 
If these gamma rays escape from the source, they are not absorbed during their propagation to the Earth, and we should observe gamma rays of GeV--TeV energies.
However, the cosmic neutrino background intensity at $\sim 10$ TeV is higher than the cosmic gamma-ray background intensity at $\sim100$ GeV, which implies that the cosmic neutrino source should be opaque to gamma rays of $\gtrsim100$ GeV \citep{Murase2016,Kimura2021_llagn}. 
The same arguments can be applied to the gamma-ray and neutrino fluxes from NGC1068, and the neutrino emission region needs to be opaque to gamma rays of $\gtrsim100$ MeV \citep{Murase2020,Inoue2020}.

In our models M2 and M3, gamma rays in $\gtrsim {\rm MeV}$ are significantly suppressed by $\gamma\gamma$ annihilation, and thus, these models are consistent with the gamma-ray and neutrino data for cosmic high-energy backgrounds and NGC 1068. The typical energy of escaping gamma-rays should be around MeV energies, and future MeV telescopes will be useful to probe the hadron-induced electromagnetic cascade emission from these systems.

\subsection{Caveats}

In this section, we discuss caveats in our model. 
First, we fixed the values of the parameters describing the AGN disks and the sBHs. 
However, in reality, these parameter should vary from source to source, affecting their contributions to the background. 
Additionally, we have not taken into account the growth of sBHs due to gas accretion (Paper~I, Section $\S~\ref{sec:evolution}$) and mergers \citep{Tagawa20_MassGap}, which is a promising pathway for sBH-sBH mergers reported by LIGO/Virgo/KAGRA \citep{LIGO20_O3_Catalog}. 
The evolution of the sBH mass influences the jet power ($L_{\rm j}$) and $T_{\rm vari}$ (the dependence of the spectral energy distribution on these quantities is presented in Appendix~\ref{sec:parameter_dependence}). 
More precise estimates for the background intensities considering these effects are desired to be conducted in the future.

Meanwhile, in our model, the AGN disk's role was considered to be just to feed gas to sBHs -- emission from shocks emerging from the AGN disk or from the broad line regions is not considered unlike paper~II, 
\citet[][paper~III]{Tagawa2023b}, or \citet{Wang2021_TZW}. 
In this paper we considered persistent high-energy emission, 
while in paper~II and paper~III, respectively, we considered transient breakout emission from shocks emerging due to collision between AGN disk gas and jets produced 
from merger remnants and solitary sBHs. 
Note that we estimated that the transient emission from shocks emerging by collision between the jets from sBHs and AGN disks cannot significantly contribute to the gamma-ray and neutrino background intensities.

In model~M1, a high Lorentz factor of 200 is assumed for fast shells (Appendix~\ref{sec:properties_jets}). On the other hand, it is unclear whether the shells in the jets can be accelerated to such a high Lorentz factor via the Blandford-Znajek process \citep[e.g.][]{Xiong2014}. 
Additionally, we assume a high contrast in the Lorentz factors between fast and slow shells (e.g. 200 and 20 in model~M1). 
However, it is unclear whether such a high contrast is commonly realized \citep[e.g.][]{Curd2022}. If either a high value or a high contrast in the Lorentz factor is not achieved, 
the dissipation rate of the kinetic energy is correspondingly overestimated in our model. 
Also, note that the equations in Appendix~\ref{sec:properties_jets} and \ref{sec:nonthermal} are approximated 
assuming that the Lorentz factor of the shocked gas is much higher than 1. 
If the Lorentz factor of the shocked gas is close to 1, the properties of electromagnetic emission and the jet structure need to be appropriately modified. 
We note that the structures of the jets and their dissipation are simply prescribed in this study. 
To quantitatively discuss the dissipation of the jets and their emission, detailed numerical simulations would be required.

The neutrino spectrum in our model might be affected by a few processes that is not included in our calculations.
First, we have not considered suppression of neutrino production 
due to the Bethe-Heitler process, which is determined by photon spectra.
The photon spectra is also modified if we take into account the electromagnetic cascade emission initiated by two-photon interactions and hadronic processes. 
In models~M2 and M3, such emission likely produces flatter spectral energy distribution below the energy limited by $\gamma\gamma$ annihilation. 
We confirmed by performing numerical computations\footnote{We modify the codes used in \cite{Kimura2019,Kimura2020_MAD_HEE} to match the physical condition in the current model. } that the suppression is not significant in our models because of the flat photon spectral produced by the cascade emission. 
The muon cooling is also not taken into account, which may reduce a neutrino flux by a factor of 2. In model M3, this process is important at $E_\nu \sim 1-10~{\rm TeV}$.

\subsection{Evolution of sBHs}

\label{sec:evolution}

Here, we discuss how much sBHs grow by accretion and their possible fate (see also \citealt{Levin2007,McKernan12}), 
as their growth is almost inevitable if they are to explain the large power required to produce the various background intensities discussed in this paper. 
In the fiducial model, the jet power is set to $2\times 10^{42}~{\rm erg/s}$, 
corresponding to the accretion rate of $3\times 10^{-4}~\Msun/{\rm yr}(\eta_{\rm j}/0.1)^{-1}$. 
Assuming that each sBH accretes for $\sim 4~{\rm Myr}$ per AGN phase and the accretion rate is constant, 
the mass of an sBH increases to $m \sim 10^3~\Msun$. 
The timescale for these (modestly) intermediate-mass BHs to migrate and inspiral to the central SMBH only by gaseous and GW torques is longer than the active AGN phase \citep{Tagawa2022_BHFeedback}, but we consider migration by stellar torques during longer quiescent phases.
Since the relaxation timescale of the nuclear star cluster is $t_{\rm rel}\lesssim 10^{11}~{\rm yr}$ \citep{Merritt2010,Kocsis2011}, 
and the migration timescale 
of an sBH with the mass $m$ is roughly given by $\sim t_{\rm rel}m_{\rm ave}/m \sim 10^8~{\rm yr} (t_{\rm rel}/10^{11}~{\rm yr})[(m/m_{\rm ave})/10^3]^{-1}$, where $m_{\rm ave}$ is the average mass of stars in a nuclear star cluster, 
the grown BHs migrate towards and accrete onto the central SMBH during quiescent phases, whose duration is roughly given as $\sim t_{\rm H}/N_{\rm AGN,H}\sim 10^9~{\rm yr}(t_{\rm H}/10^{10}~{\rm yr})(N_{\rm AGN,H}/10)^{-1}$, where 
$N_{\rm AGN,H}$ is the number of AGN phases per galaxy during the Hubble timescale, $t_{\rm H}$. 
During the Salpeter timescale of $\sim 40~{\rm Myr}$, 
inspirals of a hundred sBHs with the mass $10^3~\Msun$ enhances the SMBH mass ($10^6~\Msun$ in the fiducial model) by $\sim 10\%$. 
Thus, if the high-energy phenomena are caused by sBHs in AGN disks 
and the number of sBHs in an AGN disk is not significantly underestimated in our model, 
sBHs grow to intermediate-mass BHs, and then, they merge with SMBHs during quiescent phases, and can be observed as intermediate-mass ratio inspirals (IMRIs) by the Laser Interferometer Space Antenna \citep{Amaro-Seoane2022}.

\section{Conclusions}

In this paper, we considered high-energy EM, neutrino, and cosmic-ray emissions arising from BZ jets launched from rapidly accreting and spinning sBHs embedded in AGN disks. 
Our main results are summarized as follows:

\begin{enumerate}

\item 
GeV gamma-ray emission from nearby radio-quiet Seyfert galaxies can be explained by emission from the jets launched from sBHs if the Lorentz factor of the jets is high ($\Gamma_{\rm j}\gtrsim 30$). 
This model can contribute 
$\gtrsim 50\%$ 
of the gamma-ray emission from NGC1068 and $\sim10-20$\% of the background gamma-ray intensity in the $\sim {\rm MeV}$--$\rm GeV$ bands.

\item 
For low $\Gamma_{\rm j}$ ($\lesssim 4$), 
neutrino emission from the jets can explain the cosmic neutrino background intensity at neutrino energies $\lesssim 10^6~{\rm GeV}$.  The background in the $\sim {\rm MeV}$ bands can also be reproduced in this model if the efficiency of magnetic field amplification ($\epsilon_{\rm B}$) is low. In this case, we predict that future gamma-ray telescopes in $\sim {\rm MeV}$ bands can detect MeV gamma-rays from nearby radio-quiet AGNs, providing a test of our model.

\item 
Neutrino emission from NGC1068 can be explained by the jet model with moderate $\Gamma_{\rm j}$ and high $\epsilon_{\rm B}$, 
$\epsilon_{\rm CR}$, and $L_{\rm j}$. 
With this parameter set, jets by sBHs in AGNs can accelerate protons up to energies of $\sim 10^{19}~{\rm eV}$ 
and account for the observed intensity of UHECRs 
at these energies. 

\item 
If $\epsilon_{\rm B}$ and $\epsilon_{\rm CR}$ have a broad distribution from source to source, 
our model can simultaneously explain the neutrino flux from NGC1068 as well as the background intensities of gamma rays in $\sim {\rm MeV}$ bands, neutrinos in $\lesssim {\rm PeV}$ bands, and UHECRs, for the first time.

\end{enumerate}

While AGN are known to have relativistic jets driven by their central SMBH, our results in this paper suggest that the population of stellar-mass BHs, embedded in the accretion disk fueling the central SMBH, can collectively produce similar phenomena with energetics 
similar to 
that from the SMBH due to the larger number of jets from sBHs. 
We found that the jets from sBHs in AGN disks can explain various phenomena, 
which are observed to be unrelated to jets from SMBHs. 
Our scenario is consistent with the observations that the background neutrinos \citep{Murase2016_Neutrino} and UHECR \citep{Takami2016} emission are produced by the sources with a high local density.

\acknowledgments

We thank Kohta Murase for useful comments. 
This work was financially supported 
by Japan Society for the Promotion of Science (JSPS) KAKENHI 
grant Number JP21J00794 (HT) and 22K14028 (SSK). 
S.S.K. acknowledges the support by the Tohoku Initiative for Fostering Global Researchers for Interdisciplinary Sciences (TI-FRIS) of MEXT's Strategic Professional Development Program for Young Researchers.
Z.H. was supported by NASA grant NNX15AB19G and NSF grants AST-2006176 and AST-1715661.

\appendix

\section{Emissions}

We here describe the properties of the internal shocks in the jets 
and non-thermal emission produced from them.

\subsection{Accretion onto sBHs}
\label{sec:accretion_remnants}

We outline how the jet is launched from an accreting sBH embedded in an AGN disk (see Paper~I, for details). 
In the AGN disk, the gas accreting onto sBH forms a circum-sBH disk 
(CsBD). 
When the CsBD is advection dominated, a magnetically dominated state can be realized owing to the accumulation of the magnetic flux in the vicinity of the sBH \citep[e.g.][]{Meier2001,Cao2011,Kimura2021_BBH_PeV}. Even if the magnetic flux is initially weak, the outflow from the disk converts the toroidal magnetic field generated by the shear motion into a poloidal field \citep{Liska2020}. Such advection-dominated flows are expected for super-Eddington accretion rates \citep{Abramowicz1988} or low accretion rates 
\citep[e.g.][]{Narayan1994,Blandford1999}. 
In these cases, the jets from  spinning sBHs can be launched through the BZ process \citep{Blandford1977,Curd2022}.

Because super-Eddington accretion is predicted in an inner CsBD (Paper~I), 
the BZ jet is expected to be launched from rapidly accreting and spinning sBHs in AGN disks. 
In the process, the jet power ($L_{\rm j}$) is proportional to the mass accretion rate onto the sBH (${\dot M}_{\rm sBH}$) as 
\begin{align}
\label{eq:l_j}
L_{\rm j}=\eta_{\rm j}{\dot M}_{\rm sBH} c^2,
\end{align}
where $\eta_{\rm j}$ is the jet conversion efficiency to the kinetic energy, which is approximated as $\eta_{\rm j}\sim a_{\rm sBH}^2$ for a magnetically
dominated state \citep[e.g.][]{Tchekhovskoy2010,Narayan2021}, $a_{\rm sBH}$ is the dimensionless spin of the sBH, and $c$ is the speed of light. 
We assume that the accretion rate onto sBHs in the AGN disk is given by the Bondi-Hoyle-Lyttoleton (BHL) rate as used in Eq.~(1) of Paper~I. 
In the formula, the accretion rate is determined by 
the AGN disk density and temperature, and the Hill radius of the sBH.

\subsection{Properties of jets}

\label{sec:properties_jets}

To dissipate the kinetic energy of jets, 
we assume multiple shells with different Lorents factors collide with each other at some dissipation radius ($Z_{\rm diss}$), 
which is widely assumed to explain the prompt emission in gamma-ray bursts. 
The relative Lorents factor between the slower (with the Lorents factor $\Gamma_{\rm s}=(1-\beta_{\rm s}^2)^{-1/2}$) and faster shells (with the Lorents factor $\Gamma_{\rm r}=(1-\beta_{\rm r}^2)^{-1/2}$) is 
\begin{align}
    \Gamma_{\rm rel}=\Gamma_{\rm s}\Gamma_{\rm r}(1-\beta_{\rm s}\beta_{\rm r}) 
\end{align}
The relative Lorents factor $\Gamma_{\rm rel}$ is also related to 
the relative Lorents factors between the unshocked faster and slower shells in the rest frame of the shocked fluid ($\Gamma_{12}$ and $\Gamma_{34}$) as 
\begin{align}
    \Gamma_{\rm rel}=\Gamma_{\rm 12}\Gamma_{\rm 34}+\sqrt{\Gamma_{\rm 12}^2-1}\sqrt{\Gamma_{\rm 34}^2-1}, 
\end{align}
and the ratio of the number densities of the faster ($n_{\rm r}'$) and slower ($n_{\rm s}'$) shells in the fluid rest frame is related to them as 
\begin{align}
    \frac{n_{\rm r}'}{n_{\rm s}'}=
    \frac{(\Gamma_{\rm 34}-1)(4\Gamma_{\rm 34}+3)}{(\Gamma_{\rm 12}-1)(4\Gamma_{\rm 12}+3)}
\end{align}
The Lorents factor of the shocked jet in the rest frame of the sBH is approximately given as 
\begin{align}
    \Gamma_{\rm j}\approx \Gamma_{\rm r}
    (\Gamma_{\rm 12}-\sqrt{\Gamma_{\rm 12}^2-1}). 
\end{align}
If we assume $\Gamma_{\rm r}=200$, $\Gamma_{\rm s}=20$, and $\frac{n_{\rm r}'}{n_{\rm s}'}=\Gamma_{\rm s}^2/\Gamma_{\rm r}^2$, 
then
$\Gamma_{12}\sim 3.6$, 
and $\Gamma_{\rm j}\sim 28$, 
which we assume as fiducial values. 
Since the results are sensitive to $\Gamma_{\rm j}$, 
we also consider the models (M2, M3) with 
$\Gamma_{\rm r}=30$ and $\Gamma_{\rm s}=3$, in which 
$\Gamma_{12}\sim 3.7$, 
and $\Gamma_{\rm j}=4.2$.

\subsection{Non-thermal photons}

\label{sec:nonthermal}

We assume that the fraction $\epsilon_{e}$ of the kinetic energy of the shock is used to accelerate electrons in a collisionless shock. 
Here, the plasma and/or MHD instabilities are considered to amplify the magnetic field to $\epsilon_{\rm B}\lesssim 10^{-3}$--$10^{-1}$ 
and electrons are accelerated via the first-order Fermi process with the energy fraction of 
$\epsilon_{e}\lesssim 10^{-2}$--$0.3$
(from observations, e.g.,  \citealt{Waxman1999,Panaitescu2001,Frail2005,Uchiyama2007,Santana2014}, 
and theoretical studies, e.g., 
\citealt{Medvedev1999,Chang_2008,Spitkovsky2008b,Martins2009,Keshet_2009,Sironi_2013,Tomita2019}). 
Assuming the fast cooling regime \citep{Fan2008}, which is adequate for the internal shocks, 
the synchrotron luminosity is calculated as 
\begin{eqnarray}
\label{eq:l_sync}
L_{\rm syn} 
 \sim \frac{L_{\rm kin} \epsilon_{e} f_{\gamma \gamma}
 }{1+Y_{\rm SSC}},
\end{eqnarray}
\citep[e.g.][]{Fan2008}, 
where $f_{\gamma \gamma}$ is the attenuation fraction by the $\gamma \gamma$ annihilation given below, 
$Y_{\rm SSC}$ is the powers of synchrotron self-Compton scattering compared to that of synchrotron emission, and is calculated as 
$Y_{\rm SSC}=(-1+\sqrt{1+4\epsilon_{e}/\epsilon_{\rm B}})/2$ \citep{Fan2008}.

We assume that electrons are accelerated in the shock to a power-law distribution of Lorentz factor $\gamma'_{e}$ as 
$N(\gamma_{e}') d \gamma_{e}'\propto {\gamma_{e}'}^{-p} d \gamma_{e}'$ 
with a minimum ($\gamma_{\rm m}'$) and maximum Lorentz factors ($\gamma_{\rm max}'$). 
The minimum Lorentz factor $\gamma_{\rm m}'$ is 
\begin{align}
\label{eq:gamma_min}
\gamma_{\rm m}' \sim 
\epsilon_{e}
\left(\frac{p-2}{p-1}\right) \frac{m_{p}}{m_{e}}(\Gamma_{\rm 12}-1)\nonumber\\
\sim 
30~\left(\frac{\epsilon_{e}}{0.02}\right) 
\left(\frac{\Gamma_{\rm 12}-1}{2.6}\right)
\end{align}
for $p=2.5$,
where $m_{e}$ is the electron mass.

By comparing the cooling by the synchrotron radiation and the acceleration by the first-order Fermi acceleration mechanism, 
the maximum Lorentz factor of electrons is 
\begin{eqnarray}\label{eq:gamma_max}
\gamma_{\rm max}'
=\left(\frac{6\pi e}{\sigma_{\rm T} B'_{\rm j} \xi}\right)^{1/2}
\sim 1\times 10^6 ~
\xi^{-1/2}
\left(\frac{\Gamma_{\rm j}-1 
}{27}\right)^{-1/4}\nonumber\\
\left(\frac{\epsilon_{\rm B}}{0.01}\right)^{-1/4}
\left(\frac{n'_{p}}{9\times 10^{10}\,{\rm cm^{-3}}}\right)^{-1/4}
\end{eqnarray}
where $\xi$ is the parameter representing the ratio of the mean free path to the Larmor radius of electrons, which is adopted to be $\xi=1$ in this paper, 
\begin{eqnarray}
\label{eq:b_sf}
B'_{\rm j}=(8\pi \epsilon_{\rm B} e'_{\rm j})^{1/2}
\sim 1\times 10^4~{\rm G}~
\left(\frac{\Gamma_{\rm 12}-1 
}{2.6}\right)^{1/2}%
\left(\frac{\epsilon_{\rm B}}{0.01}\right)^{1/2}\nonumber\\
\left(\frac{n'_{p}}{9\times 10^{10}\,{\rm cm^{-3}}}\right)^{1/2},
\end{eqnarray}
is the magnetic field,
\begin{eqnarray}
\label{eq:e_sf}
e'_{\rm j}
=(\Gamma_{\rm 12}-1) n'_{p} m_{p} c^2
\end{eqnarray}
is the internal energy density of the shocked jet, and $n'_{p}$ is the proton number density of the shocked jet.

In the fiducial model, 
non-thermal emission is characterized by fast cooling regimes \citep[e.g.][]{Sari1998,Fan2008}. 
The cooling timescale for electrons with $\gamma_{\rm{m}}'$ is 
\begin{eqnarray}
\label{eq:t_minc}
t_{\rm c} (\gamma_{\rm{m}}')
\sim 0.009\,{\rm{s}} ~
\left(\frac{\epsilon_{\rm{B}}}{0.01}\right)^{-1}
\left(\frac{\gamma_{\rm 12}-1}{2.6}\right)^{-1}
 \nonumber\\
\left(\frac{n'_{p}}{9\times 10^{10}\,{\rm cm^{-3}}}\right)^{-1}
\left(\frac{\gamma_{\rm{m}}'}{30}\right)^{-1}
\left(\frac{\Gamma_{\rm j}}{30}\right)^{-1}. 
\end{eqnarray}
From $t_{\rm c} (\gamma_{\rm{m}}')$, 
the typical shell width of electrons with $\gamma_{\rm{m}}'$ emitting the synchrotron photons 
is approximated as 
$\Delta_{\rm shell}(\gamma_{\rm{m}}')\sim t_{\rm c} (\gamma_{\rm{m}}')c\sim 3\times10^{8}\,{\rm cm}~[t_{\rm c}(\gamma_{\rm{m}}')/0.009~{\rm{s}}]$.

The Lorentz factor at which self-absorption becomes effective is 
\begin{align}
\label{eq:gamma_a}
\gamma_{\rm a}'=\gamma_{\rm m}'\times 
   \left(\tau_{\rm q}C_{q+1}\right)^{1/(q+4)} 
\end{align}
 \citep{Rybicki1979,Fouka2009,Fouka2011}, 
where 
\begin{eqnarray}
\label{eq:tau_a}
\tau_q=
\frac{\pi}{3\sqrt{2}}
\frac{(q^2+q-2)\gamma_{\rm 1}'^{-5}}{1-(\gamma_{\rm 2}'/\gamma_{\rm 1}')^{-q+1}}
\frac{e n'_{\rm p} 
\Delta_{\rm shell}(\gamma_{\rm{a}}')
}{\Gamma_{\rm j}B'_{\rm j}}, 
\end{eqnarray}
$\gamma_2$ and $\gamma_1$ are the maximum and minimum Lorentz factors of non-thermal electrons with the power-law index of $q$, respectively, 
\begin{eqnarray}
\label{eq:cp}
C_q=\frac{2^{(q+1)/2}}{q+1}
\Gamma\left(\frac{q}{4}-\frac{1}{12}\right)
\Gamma\left(\frac{q}{4}+\frac{19}{12}\right), 
\end{eqnarray}
$\Gamma$ is the Gamma function, 
and we assume that electrons are randomly oriented in the frame of the shocked fluid. 
We set 
$\gamma_2'=\gamma_{\rm max}'$, $\gamma_1'=\gamma_{\rm m}'$, and $q=p_e$ for $\gamma_{\rm m}'<\gamma_{\rm a}'$, 
and 
$\gamma_2'=\gamma_{\rm m}'$, 
$\gamma_1'=\gamma_{\rm a}'$, 
and $q=2$ for $\gamma_{\rm a}'<\gamma_{\rm m}'$.

Using the variables derived above, 
we assume the spectra shapes for synchrotron emission and synchrotron-self Compton scattering as prescribed in Paper~II, 
while we additionally take into account the bolometric correction as prescribed by $R_p$ in Eq.~\eqref{eq:e_diff_cr}. 
The spectra shapes for models M1--M3 are presented in Fig.~\ref{fig:l_j1}. 
In the fiducial model, we do not consider the second order inverse Compton scattering 
as the Klein-Nishina effect is effective.

When the optical depth to the $\gamma \gamma$ annihilation ($\tau_{\gamma \gamma}$) exceeds $\gtrsim 1$, the high energy photons with $h\nu'>m_e c^2=511~{\rm keV}$ can be absorbed, 
where $h$ is the Plank constant, and $\nu$ is the frequency of photon. 
To take into account $\gamma \gamma$ attenuation, we refer to Eq.~(6) of \citet{Kimura2022_NeutrinoGRB} for $\tau_{\gamma \gamma}$. 
We assume that the attenuation factor by the $\gamma \gamma$ annihilation is given by 
$f_{\gamma \gamma}(\nu)={\rm exp}[1-\tau_{\gamma \gamma} (h\nu)]/\tau_{\gamma \gamma}(h\nu)$. 
The peak luminosity for attenuating photons is assumed to be Eq.~\eqref{eq:l_sync}, and the peak frequency to be the maximum of the minimum and absorption frequencies of the synchrotron radiation.

\subsection{Pion production efficiency}
\label{sec:neutrino_production}

We calculate the efficiency of the $p\gamma$ reaction ($f_{p\gamma}$) and the pion cooling suppression factor ($f_{\pi,sup}$) referring to Eq.~(28) and two asymptotes stated below Eq.~(33) of \citet{Kimura2022_NeutrinoGRB}, respectively. 
Comparing the dynamical timescale ($t'_{\rm dyn}$) to the $pp$ cooling timescale ($t'_{pp}$), 
the efficiency of $pp$ interaction is calculated as 
\begin{eqnarray}
\label{eq:fpp_jet}
f_{pp} 
\approx n'_{p} \kappa_{p}\sigma_{pp} Z_{\rm diss}/\Gamma_j \nonumber\\
\sim 3\times 10^{-6} 
\left(\frac{L_{\rm j}/p_\theta}{1\times 10^{44}~{\rm erg}}\right)\nonumber\\
\left(\frac{\Gamma_j}{30}\right)^{-3}
\left(\frac{Z_{\rm diss}}{5\times 10^{10}~{\rm cm}}\right)^{-1}
\end{eqnarray}
\citep[e.g.][]{Murase2014}, 
where 
\begin{align}  n'_p=L_{\rm j}/4\pi \Gamma_j^2 p_\theta Z_{\rm diss}^2 m_p c^3 
\end{align} 
is the proton number density in the jet, 
$\kappa_{p}\approx 0.5$ is the proton inelasticity, and $\sigma_{pp}\approx 4\times 10^{-26}~{\rm cm}^{2}$ is the cross section of the $pp$ interactions at 
$\sim 10$--$100~{\rm TeV}$.

To derive the efficiency of $p\gamma$ interaction, we adopt the delta function approximation. 
We assume that most interactions occur via $\Delta$-resonance process at the photon energy around the resonance peak  (${\bar \epsilon}_{\rm pk}\simeq 0.3~{\rm GeV}$), and 
approximate the cross section and inelasticity to be 
\begin{align} \sigma_{p\gamma}\kappa_{p\gamma}\simeq \sigma_{\Delta}\kappa_{\Delta} \Delta{\bar \epsilon}_{\rm pk}\delta({\bar \epsilon}_{\gamma}-{\bar \epsilon}_{\rm pk})\end{align}
where $\sigma_{\Delta}\sim 5\times 10^{-28}~{\rm cm^{-2}}$, $\kappa_{\Delta}\simeq 0.2$, and ${\bar \epsilon}_{\rm pk}\simeq 0.3~{\rm GeV}$ are the cross section, inelasticity, and the photon energy at the resonance peak, $\Delta {\bar \epsilon}_{\rm pk}\sim 0.2~{\rm GeV}$ is the peak width, $\delta(x)$ is the Dirac delta function, and ${\bar \epsilon}_{\gamma}$ is the photon energy in the proton rest frame. 
For $E_{p}<E_{p, \rm br}$, 
where 
\begin{align}E_{p, \rm br}=\frac{\Gamma_j^2 
{\bar \epsilon}_{\rm pk} m_p c^2}{2 E_{\gamma, \rm br}}, 
\end{align}
$E_{\gamma, \rm br}$ is the peak photon energy in the BH rest frame, 
the fraction of cosmic-ray protons producing pions through the photomeson production is approximated as \begin{align}\label{eq:f_pgamma}
f_{p\gamma}\approx \frac{\sigma_{\Delta} \kappa_{\Delta} \Delta{\bar \epsilon}_{\rm pk}L_{\gamma,\rm iso,br}}{2 \pi c {\bar \epsilon}_{\rm pk} \Gamma_j^2 Z_{\rm diss} E_{\gamma, \rm br} (3-s)}
(E_{p}/E_{p, \rm br})^{1-s}
\end{align}
\citep[e.g.][]{Kimura2022_NeutrinoGRB}, 
where $L_{\gamma,\rm iso,br}$ is the isotropic synchrotron luminosity at $E_\gamma=E_{\gamma, \rm br}$, 
and $s$ is the power-law slope for the synchrotron spectrum for $E_\gamma>E_{\gamma, \rm br}$. 
In model~M2, \begin{align}\label{eq:f_pgamma_num}
f_{p\gamma}\sim  7 \left( \frac{L_{\gamma,\rm iso,br}}{3\times 10^{41}\,{\rm erg/s}}\right)
\left( \frac{Z_{\rm diss}}{10^{9}\,{\rm cm}}\right)^{-1}\nonumber\\
\left( \frac{\Gamma_{\rm j}}{4}\right)^{-3}
\left( \frac{E_{\gamma, \rm br}}{200~{\rm eV}}\right)^{-1}
\end{align}
at $E_p=E_{p,\rm br}$ and $E_{p, \rm br}\sim 3\times 10^{6}~{\rm GeV}(\Gamma_j/4)^2(E_{\gamma,\rm br}/200~{\rm eV})^{-1}$. 
Since the delta function approximation cannot predict $f_{p\gamma}$ above $E_{p, \rm br}$, we do not present the neutrino luminosities in these regions in Fig.~\ref{fig:ephi_j1}. 
This approximation does not affect the detectability of neutrinos as the pion cooling suppresses the neutrino fluence in the high-energy ranges. 
In models~M1--M3, 
the $pp$ reaction dominates the neutrino reaction in lower energy ranges, while 
the ${p\gamma}$ reaction dominates over the $pp$ reaction in middle energy ranges of $10^4~{\rm GeV}\lesssim E_\nu\lesssim 10^6~{\rm GeV}$ (Fig.~\ref{fig:ephi_j1}).

\begin{figure}
\begin{center}
\includegraphics[width=85mm]{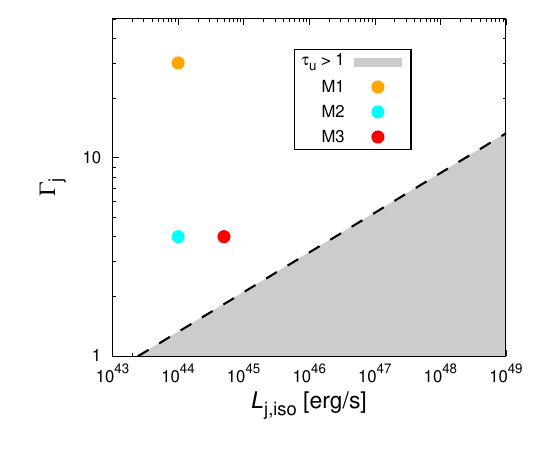}
\caption{
The parameter range for efficient particle acceleration on $\Gamma_{\rm j}$--$L_{\rm j,iso}$ plane for $\Gamma_{\rm 12}=4$ and $T_{\rm vari}=10^{-3}~{\rm s}$. 
The parameters adopted for models~M1, M2, and M3 are presented by orange, cyan, and red circles, respectively. 
}
\label{fig:tau_lgamma}
\end{center}
\end{figure}

\subsection{Parameter space for particle acceleration}
\label{sec:parameter_space}

Here we discuss the parameter space in which electrons and protons can be efficiently accelerated in internal shocks of the jets. 
If the upstream gas is optically thick to incident photons, 
the shocks are mediated by interactions with the photons \citep[e.g.][]{Ito2020}. In this case, the velocity jump at the shock is gradual, and particle acceleration should be inefficient. 
The optical depth of the upstream shell is estimated as 
\begin{align}
\tau_{\rm u}=n'_{\rm r}\sigma_{\rm T}\frac{Z_{\rm diss}}{\Gamma_{\rm r}}\nonumber\\
=\frac{L_{\rm j,iso}\sigma_{\rm T}}
{4\pi \Gamma_{\rm r}^3 Z_{\rm diss}m_p c^3}\nonumber\\
\simeq 0.004
\left(\frac{L_{\rm j,iso}}{10^{44}~{\rm erg/s}}\right)
\left(\frac{\Gamma_{\rm 12}}{4}\right)^{-3}
\left(\frac{\Gamma_{\rm j}}{4}\right)^{-5}
\left(\frac{T_{\rm vari}}{10^{-3}~{\rm s}}\right)^{-1}
\end{align}
where
$\sigma_{\rm T}$ is the Thomson scattering cross section, and we used the approximation of $\Gamma_{\rm r}\sim 2\Gamma_{12}\Gamma_{\rm j}$ in the last transformation. 
Hence, in the models we investigated (M1--M3), internal shocks are collisionless and 
particles can be efficiently accelerated (see Fig.~\ref{fig:tau_lgamma}).

\subsection{Correction for redshift evolution}
\label{sec:redshift}

We here discuss the correction factor ($f_{z}$) 
due to the redshift evolution of the number density of sources \citep{Waxman1998}, 
which needs to be taken into account to estimate the diffuse background intensity. 
For instance, $f_z \sim 3$ and $f_z\sim0.6$, 
when the number density of sources is proportional to $(z+1)^3$ and $(z+1)^0$, respectively.
The value of $f_z$ in our scenario should depend on two things: the number of sBHs captured by AGN disks and the typical accretion rate onto the sBHs. 
The accretion rate onto sBHs at a fixed radius is higher for highly-accreting AGNs, but sBHs around heavier SMBHs tend to be captured in outer regions, where the accretion rate onto sBHs is lower. These two effects likely compensate each other, and thus, we assume that the accretion rate onto sBHs is independent of the AGN luminosity.
On the other hand, the number of sBHs is higher for high-luminosity AGNs ($L_X\sim 10^{43}~{\rm erg/s}$) \citep{Tagawa20_MassGap}, whose number density strongly depends on redshift, $\sim(1+z)^3-(1+z)^4$ \citep{Ueda2014}. 
If we instead assume the number of sBHs is independent of the AGN luminosity, 
sBHs in lower-luminosity AGNs should also provide a significant contribution, whose number density is almost independent of redshift, $\sim(1+z)^0$ \citep{Ueda2014}. Thus, the source evolution in our scenario should be between $(1+z)^3$ and $(1+z)^0$. 
In this study, we adopt $f_z = 2$.

\begin{figure*}
\begin{center}
\includegraphics[width=165mm]{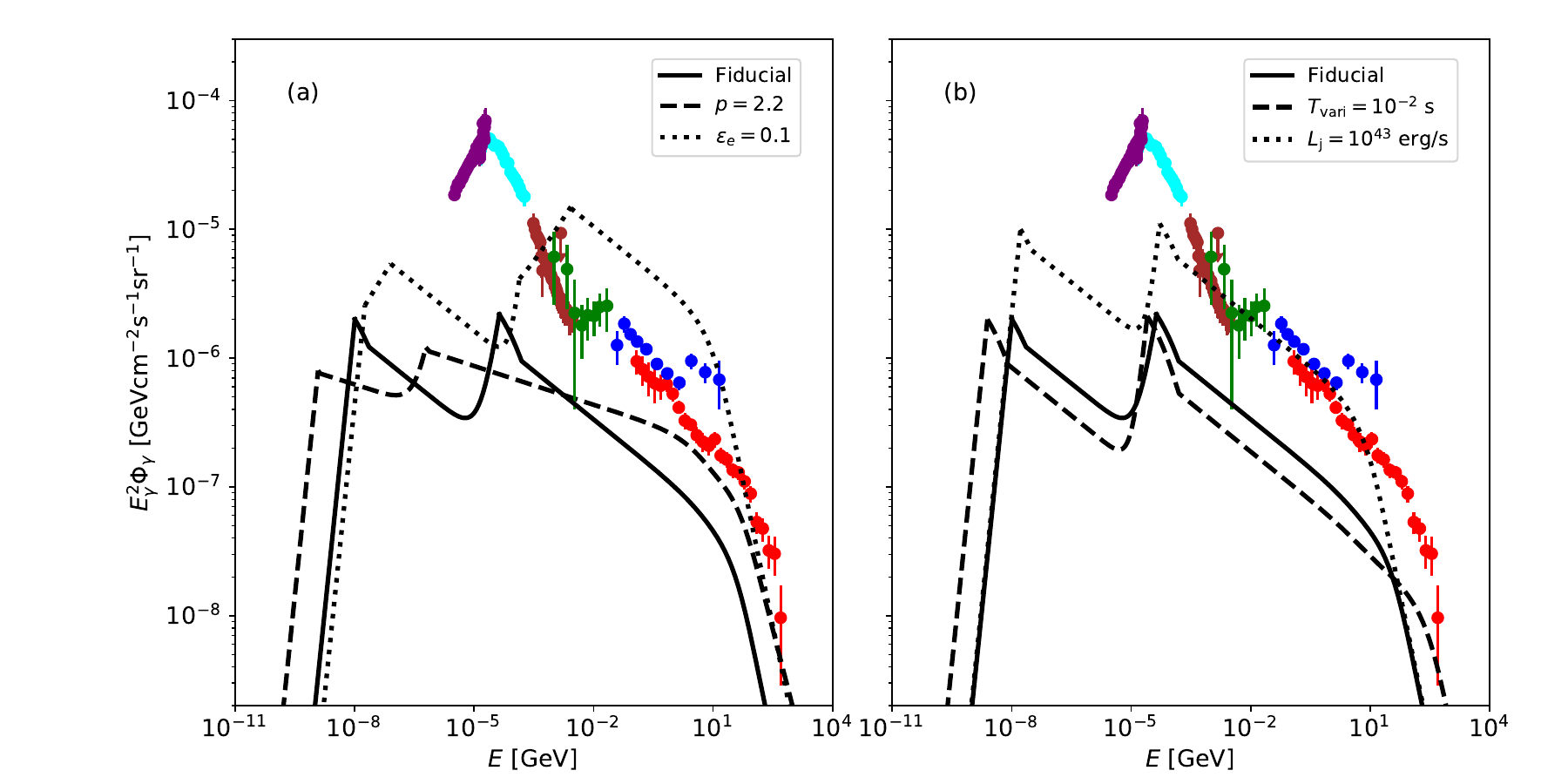}
\caption{
Same as Fig.~\ref{fig:gamma_intensity}, but with parameters varied from the fiducial model (model~M1) as indicated in the legend. 
}
\label{fig:gamma_intensity_para}
\end{center}
\end{figure*}

\begin{figure}
\begin{center}
\includegraphics[width=85mm]{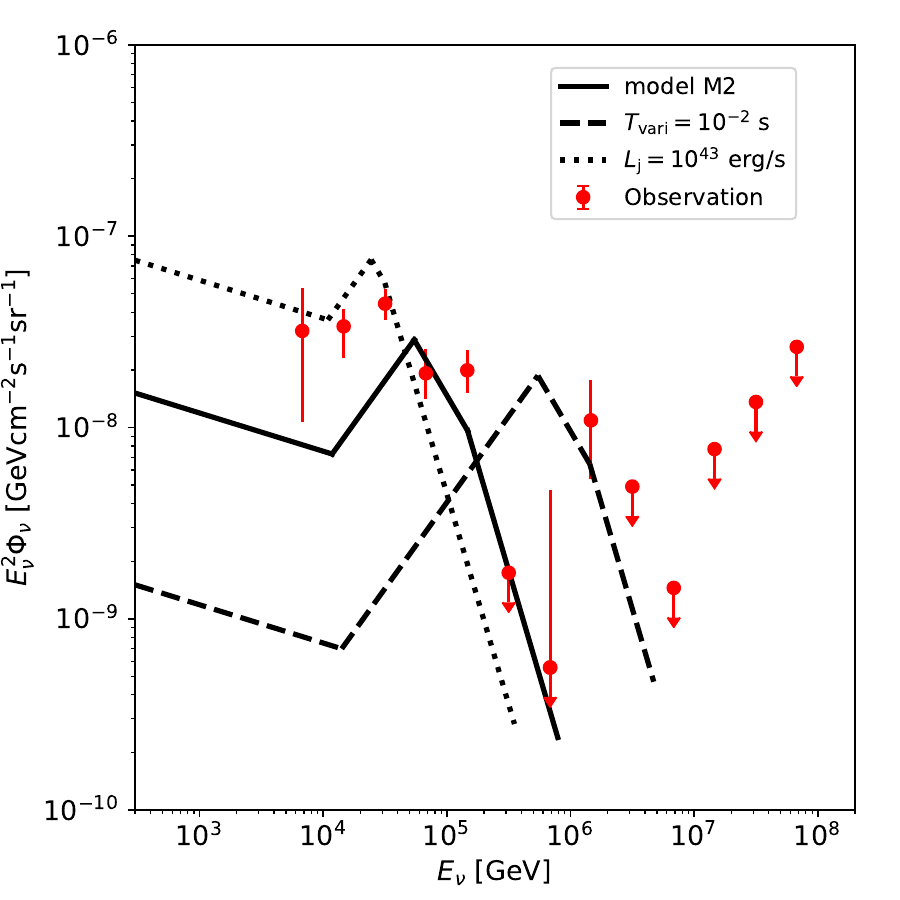}
\caption{
Same as Fig.~\ref{fig:ephi_j1}, but with parameters varied from model~M2 as indicated in the legend. 
}
\label{fig:ephi_j1_para}
\end{center}
\end{figure}

\section{Parameter dependence}
\label{sec:parameter_dependence}

In this section, we present the parameter dependence of the contribution of emission from internal shocks in jets to the gamma-ray (Fig.~\ref{fig:gamma_intensity_para}) and neutrino background intensities (Fig.~\ref{fig:ephi_j1_para}).

Fig.~\ref{fig:gamma_intensity_para} shows the dependence of 
the gamma-ray flux from internal shocks on the parameters, $p$, $\epsilon_e$, 
$T_{\rm vari}$, and $L_{\rm j}$. 
All of these parameters influence the minimum synchrotron frequency ($\nu_{\rm m}=\gamma_{\rm m}^2\nu_{\rm syn}$, where $\nu_{\rm syn}$ is the synchrotron frequency). 
This is obvious that 
$\gamma_{\rm m}$ depends on 
$\epsilon_e$ and $p$ (Eq.~\ref{eq:gamma_min}), while $\nu_{\rm syn}$ depends on 
$T_{\rm vari}$ and $L_{\rm j}$ as $\nu_{\rm syn}\propto B_{\rm j}'\propto n_p'^{1/2}\propto L_{\rm j}^{1/2}T_{\rm vari}^{-1}$. 
In addition, $p$ influences 
the power-law slope above $\nu_{\rm m}$ (dashed black line in Fig.~\ref{fig:gamma_intensity_para}a), 
and $\epsilon_e$ and $L_{\rm j}$ influence the total radiation energy by non-thermal emission (dotted black lines in Fig.~\ref{fig:gamma_intensity_para}). 
Note that $n_{\rm rotBH,acc}$ also linearly affects the intensity ($E_\gamma^2 \Phi_\gamma \propto n_{\rm rotBH,acc}$). 
Thus, the gamma-ray intensity depends on various parameters. 
Nevertheless, gamma-ray emission is not eliminated by these parameters, unlike the model with different $\Gamma_{\rm j}$ as shown in the main manuscript (Fig.~\ref{fig:gamma_intensity}).  

Fig.~\ref{fig:ephi_j1_para} shows the dependence of the neutrino flux from the internal shocks on several parameters. 
The neutrino intensity in low energy ranges is low and high for the long-$T_{\rm vari}$ or high-$L_{\rm j}$ models (dotted black and solid orange lines) since $f_{pp}$ is proportional to the high jet density ($f_{pp}\propto n_p'\propto L_{\rm j}T_{\rm vari}^{-2}$). Also, 
for long $T_{\rm vari}$ or low $L_{\rm j}$, 
cooling by synchrotron radiation during pion phases is inefficient, due to weak magnetic fields in shocks, 
and then the neutrino intensities are high in high energy ranges. 
Similar to the gamma-ray intensity, $n_{\rm rotBH,acc}$ linearly influences the neutrino intensity ($E_\nu^2 \Phi_\nu \propto n_{\rm rotBH,acc}$).

\bibliographystyle{aasjournal}
\bibliography{agn_bhm}

\end{document}